 \documentstyle[12pt]{article}
\begin{document}
 \newcommand{\be}[1]{\begin{equation}\label{#1}}
 \newcommand{\ee}{\end{equation}}
 \newcommand{\beqn}[1]{\begin{eqnarray}\label{#1}}
 \newcommand{\eeqn}{\end{eqnarray}}
\newcommand{\mat}[4]{\left(\begin{array}{cc}{#1}&{#2}\\{#3}&{#4}\end{array}
\right)}
 \newcommand{\matr}[9]{\left(\begin{array}{ccc}{#1}&{#2}&{#3}\\{#4}&{#5}&{#6}\\
{#7}&{#8}&{#9}\end{array}\right)}
 \newcommand{\eps}{\varepsilon}
 \newcommand{\Ga}{\Gamma}
\newcommand{\s}{\sigma}
\newcommand{\D}{\Delta}
 \newcommand{\la}{\lambda}
\newcommand{\ov}{\overline}
\newcommand{\mucirc}{\stackrel{\circ}{\mu}}
\newcommand{\mast}{\stackrel{\ast}{m}}
\newcommand{\meps}{\stackrel{\circ}{\eps}}
\newcommand{\mcirc}{\stackrel{\circ}{m}}
\newcommand{\mcir}{\stackrel{\circ}{M}}
\newcommand{\geqsim}{\stackrel{>}{\sim}}
\renewcommand{\thefootnote}{\fnsymbol{footnote}}

\begin{titlepage}
\begin{flushright}

INPP-UVA-95/09\\
November, 1995
\end{flushright}
\vspace{10mm}

 \begin{center}

 {\Large \bf Solution to the Strong CP Problem: \\
Supersymmetry with Parity}\\

\vspace{1.3cm}
{\large Ravi Kuchimanchi }
\\ [5mm]
{\em Department of Physics,
University of Virginia, Charlottesville, VA 22901, U.S.A.}\\
\end{center}

\vspace{2mm}
\begin{abstract}


We find that supersymmetry with parity can solve the strong CP problem
in many cases including the interesting cases of having the
minimal supersymmetric standard model or some of its extensions
below the Planck/GUT/intermediate scales, as well as for the
case where we have a low-energy SUSY left-right model.
\end{abstract}

\end{titlepage}

\renewcommand{\thefootnote}{\arabic{footnote})}
\setcounter{footnote}{0}


The smallest dimensionless parameter in the standard model is the strong
CP phase, $\bar{\theta} = \theta + Arg Det (M) $ where $\theta / 32 \pi^{2}$
is the cofficient of the $F \tilde{F}$ term in the QCD lagrangian and $M$ is
the quark mass matrix.   Experimental bounds on the electric dipole
moment of the neutron imply that $\bar{\theta} \le 10^{-9}$.  There
is no symmetry by which $\bar{\theta}$ can be made naturally small (or zero)
at the level of the standard model, and this has been called the strong CP
problem.

Two elegant solutions have been proposed.  If the up quark were massless
or if there were a $U(1)$ Peccei-Quinn (PQ) symmetry~\cite{pec}, then
$\bar{\theta}$
can be rotated away.  However the up quark seems to be massive, and the
PQ symmetry leads to an axion which is severely constrained by experiments.
Other
solutions like spontaneous CP violation or  the Nelson-Barr
mechanism~\cite{nel84},
require heavy quarks. Solutions based on
spontaneous P violation have   so far required mirror
families~\cite{sen91} or CP symmetry as well~\cite{beg78}.    While none of
the existing solutions are completely
ruled out, nevertheless a solution to the strong CP problem
continues to occupy our minds.

In the supersymmetric extension of the standard model, not only has no new
solution to the strong CP problem been found, but also
a new problem gets generated - namely the small SUSY phase
problem~\cite{bab94}.  Even
if the strong CP problem were solved, for example by the PQ symmetry,
direct contributions to the dipole-moment of the neutron constrain
many other CP violating phases in the theory which could be apriori of the
order 1.  Also the Nelson-Barr mechansim does not seem to generalize to
the supersymmetric extension of the standard model even with universal
soft SUSY breaking terms at the Planck/GUT scales~\cite{bar93}. Thus
solutions to the strong CP problem based
on spontaneous CP violation or Nelson-Barr mechanism have not been
extended to MSSM and this is grounds for serious concern.  From
an experimental point of view, while the MSSM is very predictive on things
like the Higgs mass, it does very poorly on the important question
of additional CP phases.  At least two new independent phases in the
$A, B, \mu, m_{1/2}$ terms are expected and so far there is no
theoretical predication on their values~\cite{gar94}.

In this letter we show that if the minimal supersymmetric standard
model (MSSM)~\cite{mssm} is extended to include parity (which can then
be broken to MSSM at any scale between $M_{SUSY}$ and $M_{Pl}$ )
a new solution to the strong CP problem is obtained. Further the small
SUSY phase problem is also automatically solved. An important
prediction emerges that there are no phases in the ratio of the
Higgs vevs, $\mu, B, A,$ or $m_{1/2}$ of MSSM.  We also study the
solution for low-energy SUSY left-right symmetric model.
\vskip0.2in

\noindent{\underline{Strong CP problem with Parity:}} To be specific we
include
parity by extending the standard model to the left-right symmetric model
$SU(2)_{L} \times SU(2)_{R} \times SU(3) \times U(1)_{B-L}$~\cite{lr}.  The
matter spectrum consists of the usual quarks and leptons,
$Q_{i}(2,1,3,1/3),$
$Q^{c}_{i}(1,2,\bar{3},-1/3)$, $L_{i}(2,1,1,-1)$ and $L^{c}_{i}(1,2,1,1)$
where $i$ is the generation index and runs from 1 to 3.
One or more (indexed by $a$) bidoublet Higgs fields $\phi_{a} (2,2,1,0)$ are
introduced to break the theory down to electromagnetism.
The $\phi_{a}$ are each represented by $2 \times 2$ complex matrices
while the doublet quark and lepton fields by $2 \times 1$ column vectors.
Under parity
{\bf x} $\rightarrow$ {\bf - x}, $Q_{i} \leftrightarrow Q^{c*}_{i}$,
$L_{i} \leftrightarrow L^{c*}_{i}$ and $\phi_{a} \leftrightarrow
\phi_{a}^{\dagger}$.  Invariance under parity of the Yukawa term
$(h_{ij}^{a} Q_{i}^{T} \phi_{a} Q^{c}_{j} + h.c.)$ implies that
$h^{a}_{ij} = h^{a*}_{ji}$, (ie) the Yukawa matrix is hermitian.
The mass matrix is the product of the Yukawa matrix and the vacuum
expectation values (VEVs) $\left < \phi_{a} \right >$.  Therefore
the mass matrix will have a real determinant if we can prove that
the matrices $\left < \phi_{a} \right >$ are real.  This would then lead to
a solution of the strong CP problem since the coupling $\theta$ of the
parity odd $\theta/ 32 \pi^{2}~ F \tilde{F}$ term is zero due to parity.

$\left < \phi_{a} \right >$ are determined by minimizing the Higgs potential
and can be naturally real only if all the coupling constants involving
$\phi_{a}$ in the Higgs potential are real.  We begin by making
an important observation that terms involving only $\phi_{a}$ are of the
form $m_{ab} Tr \phi_{a}^{\dagger} \phi_{b}$, $\mu_{ab} Tr (\tau_{2}
\phi_{a}^{T} \tau_{2} \phi_{b})$, etc. (in general
traces of products of $\phi_{a}, \phi_{a}^{T}, \phi_{a}^{\dagger}$ and
$\tau_{2}$).  By comparing every term and its hermitian conjugate, it is easy
to see that invariance under P $(\phi_{a} \leftrightarrow
\phi_{a}^{\dagger})$ implies that all the constants $m_{ab}, \mu_{ab},$
etc. are real!  If we have additional gauge singlet Higgs fields
$\sigma$, such that under $P: \sigma \leftrightarrow \sigma^{\dagger}$, then
all coupling constants of terms involving $\phi_{a}$ and $\sigma$ will
also be real.

In order to break $SU(2)_{R} \times U(1)_{B-L}$ to $U(1)_{Y}$ at a
scale $M_{R}$ (which can be anywhere between $M_{W}$ and the
Planck scale $M_{Pl}$,
we need to introduce Higgs triplet or doublet fields, namely,
either $\Delta(3,1,1,2)$, $\Delta^{c}(1,3,1,-2)$ OR
$\chi(2,1,1,1)$, $\chi^{c}(1,2,1,-1)$ such that under P:
$\Delta \leftrightarrow \Delta^{c*}, \chi \leftrightarrow \chi^{c*}$
and give a VEV to the right handed fields.  There will be
coupling terms between $\phi_{a}$ and $\Delta^{c}$ or
$\chi^{c}$.  Terms of the form $\lambda \left ( \Delta^{c \dagger}
\Delta^{c} Tr (\tau_{2} \phi_{a}^{T} \tau_{2} \phi_{a}) +
\Delta^{\dagger} \Delta Tr ( \tau_{2} \phi_{a}^{\dagger} \tau_{2}
\phi_{a}^{*}) \right )$ are invariant under P, and  $\lambda$ can
be complex. We note that this complex term is the source of the
strong CP problem in left-right symmetric theory.


If there is supersymmetry, as we shall see, these terms coupling
$\Delta^{c}$ to $\phi$ with complex coupling constants
 are naturally
absent and we are led to a solution to the strong CP problem.  The
rest of the paper analyzes the tree level and loop effects of the
solution in SUSY left right models~\cite{kuc93,cve84,fra91} spontaneously
breaking to MSSM.
\vskip0.2in
\noindent {\underline {Tree level Solutions: SUSY with Parity}} \\

\noindent {\it Case 1: Minimal left-right model } -  The superpotential
for the minimal model is given by~\cite{kuc93,cve84,fra91}
\begin{equation}
W = M \ Tr \ \Delta^{c} \bar{\Delta^{c}} + M^{*} \ Tr {\Delta \bar{\Delta}}
+ \mu_{ab} \  Tr \ \tau_{2} \phi^{T}_{a} \tau_{2} \phi_{b}
\end{equation}
There is no coupling between the $\Delta^{c}$ and the $\phi_{a}$.
This is the case even for the most general soft SUSY breaking terms
since they are given by the most  general analytic cubic polynomials
in the scalar fields of the theory.
Since these have the same form as $W$ (but with arbitrary coefficients),
there is no coupling between $\Delta^{c}$ and $\phi_{a}$ in these terms
either. The D-terms only involve real gauge coupling constants.  As
explained in the previous section,
$\mu_{ab}$ and coupling constants of the quadratic
soft SUSY breaking terms involving $\phi_{a}$, are all real due to parity.
Hence {\it all} coupling constants in the Higgs potential wherever
$\phi_{a}$ occurs are real.  Thus $\left < \phi_{a} \right >$  is naturally
real and at the tree level there is no Strong CP problem.

We would like to preserve this nice feature of the minimal model
while extending to non-minimal models.  The main reasons to extend
are that we need to break the left-right symmetric theory to
MSSM at a high scale- so we {\it have} to introduce other fields.
Also as it stands this model will break $Q_{em}$ spontaneously
unless R-parity is broken by giving the sneutrino a VEV~\cite{kuc93}.
We  would like to keep R-parity unbroken, so as not
to introduce the complex phases in the lepton sector
and make the problem more complicated.   This is another reason
to consider non-minimal models. From now on we will interchangably
use $\phi$ for $\phi_a$ since the generalization to more than one doublet
is now obvious.  Also, in the following we will not explicitly write
the squark or slepton fields as their VEVs are zero.\\
\vskip0.2in
\noindent {\it Case 2: Breaking to MSSM + Singlet} -
In order to solve the $\mu$ problem~\cite{mssm}, MSSM with a singlet $\sigma$
has been considered previously in the literature.
A discrete $Z_3$ symmetry $\phi, \sigma, Q, Q^{c}, L, L^{c} \rightarrow
e^{i 2 \pi /3} \left ( \phi, \sigma, Q, Q^{c}, L, L^{c} \right )$
prevents the
direct $\mu$ - term.  We will show that SUSY left-right symmetric
theory can naturally break
to this low-energy theory with zero tree-level strong CP phase.  The most
general left-right symmetric superpotential is
\begin{eqnarray}
W = & &  M Tr \Delta^{c} \bar{\Delta^{c}} + M^{*} Tr \Delta \bar{\Delta}
+ \beta \left ( h_{\beta} Tr \Delta^{c} \bar{\Delta^{c}} + h_{\beta}^{*}
Tr \Delta \bar{\Delta} \right )  + f(\beta)  \nonumber \\
& & + h_{\sigma} \sigma  Tr \tau_{2} \phi^{T} \tau_{2} \phi
+ \lambda \sigma^{3}
\end{eqnarray}
where $f(\beta)$ is any cubic polynomial, and under $Z_{3}$,
$\Delta, \Delta^{c} \rightarrow e^{i 2 \pi / 3} ( \Delta, \Delta^{c} )$,
$\bar{\Delta}, \bar{\Delta}^{c} \rightarrow e^{- i 2 \pi / 3} (
\bar{\Delta}, \bar{\Delta}^{c} )$ and $\beta \rightarrow \beta$.
Under parity $\sigma \rightarrow \sigma^{\dagger}$ and $\beta \rightarrow
\beta^{\dagger}$.
The soft SUSY breaking terms can have  their most general form
consistent with Parity and $Z_3$.  F-terms are obtained by taking the
partial derivative of $W$ with respect to each of the fields
in the superpotential (denoted
here by $A_{i}$), so that
\begin{equation}
V_{F} = \Sigma_{i} {\left | {{\partial W} \over {\partial A_{i}}} \right
|}^2 .\end{equation}
It is easy to see that there
are solutions for $V_{F} \approx M_{SUSY}^4$ such that $\Delta^{c},
\bar{\Delta^{c}} \approx M_{R}$
and $\phi, \sigma$ are less than $M_{SUSY}$. This implies that we
can break the theory down to MSSM + singlet at a high scale $M_R$.
Once again since there
is no coupling terms between the $\Delta^{c}$ and $\phi$ fields,
the coupling constants in the Higgs potential for
 {\it all} the terms which contain  $\phi$
are real due to parity. Likewise, all coupling constants for terms involving
$\sigma$ are also real.  Thus $\left < \phi \right >$ and
$\left < \sigma \right >$ are naturally real and there is no strong
CP phase at the tree level. A point to note is that  this
model has not been
considered
in Reference~\cite{kuc93}.  Therefore the result of that paper does not
apply and there can be $Q_{em}$ conserving and parity breaking vacuua
without needing R-parity breaking.  This is because a complex $h_{\beta}$
leads to a complex VEV for $\beta$, thereby breaking parity.  The
quartic F-term in $\phi$ stabilizes the $Q_{em}$ conserving vacuum.
\vskip0.2in
\noindent{\it Case 3: Breaking to MSSM} -
We introduce singlet (they could be in general
triplet or other fields too) fields $\alpha, \beta$ and $ \gamma$.
Under Parity they go to their Hermitian conjugate fields.
The most general super potential is given by:
\begin{eqnarray}
W = & &  h_{\alpha} \alpha Tr \Delta^{c} \bar{\Delta}^{c} +
h_{\beta} \beta \Delta^{c} \bar{\Delta}^{c} + m_{\alpha} \alpha \gamma
+ m_{\beta} \beta \gamma + \nonumber \\
& & \lambda_{1} \gamma^{3} + \lambda_{2} \alpha^{3}
+ \lambda_{3} \beta^{3} + \mu_{ab} Tr \tau_{2} \phi_{a}^{T} \tau_{2} \phi_{b}
+ ( \Delta~ terms)
\end{eqnarray}
where in order to prevent couplings between $\Delta^{c}$ and $\phi$
in the F-term, we have imposed a discrete symmetry (D) such that
$\gamma, \bar{\Delta}, \bar{\Delta}^{c} \rightarrow
e^{i 2 \pi / 3} ( \gamma, \bar{\Delta}, \bar{\Delta}^{c} )$,
$ \alpha, \beta \rightarrow e^{- i 2 \pi / 3} \left ( \alpha, \beta
\right ) $ and the rest of the fields are invariant.
The singlets allow us to break the $SU(2)_R \times
U(1)_{B-L}$
symmetry at a high scale ( $M_{R}$ ) to MSSM.
This is important and can be  checked by writing out all the F-terms
obtained by differentiating $W$ with each and every field.  The crucial
point is that there
are solutions for $V_{F} = 0$, with the singlets and the right-handed
fields {\it alone} picking up VEVs. $m_{\alpha}$ and $m_{\beta}$ set
the scale for $M_{R} >> M_{SUSY}, M_{W}$.
Once again all coupling constants in terms
wherever $\phi$ occurs are real, and hence $\left < \phi \right >$ is real
and there is no strong CP problem.  In order to avoid the bound
$M_{R} \le M_{SUSY} / f$ of reference~\cite{kuc95} we can introduce
$B - L = 0$ triplet fields $\omega$, $\omega^{c}$ such that
under D: $\omega, \omega^{c} \rightarrow e^{ - i 2 \pi / 3}
\left ( \omega, \omega^{c} \right )$. This
does not change the result that $\left < \phi \right > $ is real.

This case has the advantage over case 2 since the discrete symmetries
(Parity as well as the discrete symmetry D)
can be broken at a high scale so that domain walls associated
with the breakdown of discrete symmetries can be inflated away.
In case 2 since there is a residual low-energy $Z_{3}$ symmetry this problem
exists (since this symmetry breaks only when $\phi$ picks up a VEV).

We have shown three illustrative cases where there is a natural solution to
the strong CP problem at the tree level.  Other non-minimal models
can be easily accomadated, in a similar manner.  Now we will study
the loop effects. \\
\vskip0.2in

\noindent {\underline{The Complete Solution}}
\vskip0.2in
If $M_{R} >> M_{SUSY}, M_{W}$ then the effective theory below $M_{R}$
will be SUSY $SU(2)_L \times SU(3) \times U(1)$ (and in particular in
case 3 it will be the MSSM).
The MSSM superpotential and the soft SUSY breaking terms are given
by~\cite{mssm,mar94,kan94}:
\begin{equation}
W = \mu H^{T} \bar{H} + h^{u}_{ij} Q^{T}_{i} H u^{c}_{j} +
 h^{d}_{ij} Q^{T}_{i} \bar{H} d^{c}_{j}
\end{equation}
\begin{eqnarray}
V_{S} = & & m H^{T}  \bar{H} + A^{u}_{ij} \tilde{Q}^{T}_{i} H
\tilde{u}^{c}_{j} +  A^{d}_{ij} \tilde{Q}^{T}_{i} \bar{H} \tilde{d}^{c}_{j}
\nonumber \\
& & + M_{3} \tilde{G} \tilde{G} + M_{2L}\tilde{W_L} \tilde{W_{L}} + M_{Y}
\tilde{Y} \tilde{Y} + quad. ~scalar~ masses.
\end{eqnarray}
where, $\tilde{G}$ and $\tilde{W_L}$ are the Gluino and Left-Handed
Gaugino (wino) respectively and
where the standard model Higgs doublets are denoted by $H(2,1,-1)$ and
$\bar{H}(2,1,1)$.  These doublets are the light elements of the
bidoublet fields $\phi_{a}$.
Due to parity and since we have shown that all
couplings in terms containing $\phi_{a}$ are real
(in cases 1,2 and 3 above even after $\Delta^{c}$ picks up VEV),
the following boundary conditions emerge at $M_{R}$ :
\begin{equation}
\label{eq:bc}
\mu = \mu^{*}, \  h^{u}_{ij} = h^{u*}_{ji}, \ h^{d}_{ij} = h^{d*}_{ji}, \
A^{u}_{ij} = A^{u*}_{ji}, \ A^{d}_{ij} = A^{d*}_{ji}, \
m = m^{*}, M_3 = M_3^{*}
\end{equation}
We  now use the two loop renormalization group equations~\cite{mar94}
to run the above coupling constants to the $TeV$ scale. Our results
follow.\\
\vskip0.1in
\noindent {\it Result 1: With universal soft SUSY breaking terms} -
If the soft SUSY breaking terms come from a supergravity
sector,  there are further constraints that soft SUSY breaking
terms can satisfy~\cite{kan94}.  These constraints can both be
derived from some supergravity theories, and can be
motivated by low energy flavour phenomenology. We will first consider
the most constrained MSSM that has received the maximum attention
with the following universality conditions at the SUGRA breaking
(or Planck) scale.
\begin{equation}
A^{u,d}_{ij} = A h^{u,d}_{ij}, \ \  m = B \mu, \ \  M_{3} = M_{2L} =
M_{2R} = M_{B-L} = m_{1/2}. \end{equation}
Now using equation~\ref{eq:bc} it is easy to see that $A, B$ and
$m_{1/2}$  real. The quadratic scalar
masses are also universal. Therefore the only complex phase in the
theory is the standard model  CKM matrix phase. It is easy to see
using the 2-loop renormalization group equations (RGE) of
reference~\cite{mar94} that
every coupling constant (coupling matrix) and its hermitian conjugate evolve
according to the same RGE if the above conditions are met.
Therefore Hermitian matrices remain Hermitian and real couplings
remain real.   Thus
at the weak scale the Yukawa and squark matrices are Hermitian.
The Higgs doublet coupling constants are all real, and  the Gluino,
Bino and  Wino mass terms are real.  Hence the expectation value
of $H$ and $\bar{H}$ is real, and the quark mass matrix
is Hermitian and the Strong CP phase is zero.  In this case, the loop
effects at the weak scale  will induce a negligibly
small strong CP phase consistent with $\bar{\theta} << 10^{-9}$, and
we have the solution to the strong CP problem.
Note that we have implicitly assumed that $M_R$ is approximately
equal to the SUGRA breaking scale (or $M_{Pl}$ ) and we will
relax this condition later (see Result 3). \\
\vskip0.1in
\noindent {\it Result 2: With univerality only for Gauginos} -
There are supergravity models where only some but not necessarily
all universality conditions are predicted.  Also in string theory
we may have non-universal terms~\cite{iba92}.
The only universality condition that is really needed for us is
that of gaugino phases, namely,
\begin{equation}
Arg M_{3} = Arg M_{2L} = Arg M_{2R} = Arg U(1)_{B-L} = Arg M
\end{equation}
We will not assume any other universailty condition and so the
rest of the soft SUSY breaking terms can be general.
Even in this case, and using
equation~\ref{eq:bc}, it is easy to see that the RGE
preserve the Hermitian
and real nature
of the respective coupling constants and
therefore, just as in Result 1 it follows that there is no strong CP
problem.  In addition to supergravity models already
included in the first result, such a universality condition
can be obtained from
models where the gaugino mass term ratios depend only on real numbers
like the structure constants of the gauge groups~\cite{iba92}.  It can also
happen due  to an underlying grand-unified group. \\
\vskip0.1in
\noindent {\it Result 3: Bound on Wino phase, accessible strong CP} -
Parity only relates the left and the right Wino phases but does not set them
to zero.  If instead of at $M_{Pl}$, $SU(2)_{R}$ is broken at an
intermediate scale $M_{R}$ then even if wino phases are real at
$M_{Pl}$ they will pick up a complex value from the complex terms in the
$\Delta$ sector, due to renormalization group running from $M_{Pl}$ to
$M_{R}$.    This phase will inturn give rise
to a gluino phase because while the left handed wino contributes to the
renormalization group running from $M_{R}$ to $M_{W}$,
the right handed wino does not not.  Both effects are at the two loop
level~\cite{mar94}.  Hence the gluino mass term picks up a phase of the
order ${(1 / 16\pi^{2})}^2 \times {(1 / 16\pi^{2})}^2 \times
\delta$, which is about $10^{- 9} \times \delta$ where $\delta$
are typical phases in the $\Delta$ terms.   This resultant
strong CP phase is consistent with
current experimental bounds, and at the same time is reasonably
exciting for the neutron electric dipolemoment searches.

Even if the Planck scale universality conditions on the gaugino
phases are not exact, what the above estimate implies is that
the wino phases must be within ${( 1 / 16 \pi^{2})}^{2}$ at that
scale or they will induce too large a strong CP phase.

The only exception will be if  $M_{R} < M_{SUSY}$ because in this case
we have a low energy SUSY left-right symmetric theory, and the left and
right handed wino mass term phase contributions to the Gluino
phases will cancel. Note that the illustrative models
of case 1, 2, and 3 can have $M_R \le M_{SUSY}$ \\
\vskip0.1in
\noindent{\it Result 4: Without Universality, $M_{R} \le M_{SUSY}$} -
Even if the wino phases are greater than  $10^{- 4.5}$
 ( and the soft SUSY breaking terms have their most general form), this
solution to the strong CP  problem works.  This is  because both
left and right winos are present at low energies
if $M_{R} < M_{SUSY}$, and their effects will
cancel.  However
since
$M_{R} > M_{W}$ there will be a mixing between the right wino and the
right handed Higgsinos, due to the VEVs $\left < \Delta^{c} \right >$
and $ \left < \bar{\Delta}^{c} \right > $.  However both expectation
values are required to induce a phase, and this phase will therefore be
of the order
$ \delta \times  \left < \Delta^{c} \right > \left <
\bar{\Delta}^{c} \right >
/ M_{SUSY}^{2}$, where
$\delta$ are the typical phases in $\Delta^{c}$ terms.   This must be less
than $10^{- 4.5}$ to be
consistent with the neutron edm.  This bound does not necessarily imply a
high $M_{SUSY}$, because it is possible to choose
soft SUSY breaking parameters such that one of the
$SU(2)_{R}$ triplet fields picks up a much smaller VEV than the other
$SU(2)_R$ triplet.
However these choices of
parameters
may not be consistent with some of the universality conditions.
Only in such cases (and if we are working in supergravity theory which
gives the universality condition)
this implies the bound $\delta \times M_{R}^{2} / M_{SUSY}^{2}
 \le 10^{-4.5}$.

Another very interesting possibility for low-energy left-right symmetry is
with $ \chi, \chi^{c}, \bar{\chi},
\bar{\chi}^{c}$ fields instead of the triplets, and
with universal $A$ and $B$ terms.  Once we impose R-parity, and noting
that terms like $\chi^{T} \phi \chi^{c}$ must have a real coupling
constant (due to parity), we can see that the phase in the
$\chi^{cT} \bar{\chi^{c}}$ can be rotated away.  Universal
$A$ and $B$ terms must be real due to parity.  Hence we once
again have real vacuum expectation values for the Higgs fields.
There is no problem with having to worry about $Q_{em}$ breaking minima as
the result
 of Reference~\cite{kuc93}  does not
apply to doublet fields.  In this model the majorana mass term for
the right-handed neutrino can arise in loops (or the other option is to
introduce extra R-parity odd singlets, "neutrinos", to
be able to give a large  mass to right-handed neutrinos).
But this solution requires $M_{R} \le M_{SUSY}$, since introducing
any R-parity even singlets to increase the scale of $M_{R}$ will lead to
additional complex couplings.  The only way to allow the singlets
would be to prevent the coupling term $\chi^{T} \phi \chi^{c}$
by a symmetry.
Once we do this
the rest of the analysis for $\chi^{c}$ fields is exactly similar to
that for breaking $SU(2)_{R}$via the $\Delta^{c}$ fields.
But now the Majorana mass cannot be generated
by loops, and we need to introduce extra neutrinos.  \\
\vskip0.2in
\noindent {\underline{Non-supersymmetric left-right model}}\\
It is worth noting that the term
$\lambda (\Delta^{c \dagger}\Delta^{c} \ Tr \left ( \tau_{2} \phi^{T}
\tau_{2} \phi \right ) )$ that was the source of the strong
CP problem in non-supersymmetric left-right models can be eliminated
by discrete symmetries like $ \phi_{1} \rightarrow i \phi_{1}$,
$Q^{c}_{3} \rightarrow - i Q^{c}_{3}$ etc.  We need to introduce
enough bidoublets and have enough non-trivial symmetries that
transform the quark fields in a family number dependent way, so
as to prevent all such troublesome terms, while at the same time
obtaining a consistent quark mass matrix.
These symmetries are in the low-energy theory and hence will
lead to the domain wall problem, unless soft breaking terms
(that break these symmetries, but not Parity) of dimension 2 are allowed.
In this case we can have solutions to the strong CP problem
in the non-supersymmetric version, without having to introduce
CP as a good symmetry of the Lagrangian.
\vskip0.2in
\noindent {\underline {Conclusions}} \\
We have shown that supersymmetry with parity can solve the
strong CP probem in many cases including the interesting cases
of having the
MSSM or some of its extensions below the Planck/GUT/intermediate scales, as
well as  for the case where we have a low-energy SUSY left-right model.

\vspace{4mm}
I would like to thank P.K. Kabir and P.Q. Hung for discussions.
The work was supported
 by the U.S. Department of Energy under grant No. DE-AS05-89ER40518.

\end{document}